\documentclass[seceq]{ptptex}
\usepackage{graphicx}

\def\sign{\mathop{\rm sign}\nolimits}

\newcommand{\al}{{\alpha}}
\newcommand{\bt}{{\beta}}
\newcommand{\ga}{{\gamma}}

\newcommand{\ep}{{\epsilon}}
\newcommand{\zt}{{\zeta}}

\newcommand{\sig}{{\sigma}}

\newcommand{\pd}{{\partial}}

\newcommand{\wt}{\widetilde}

\newcommand{\U}{{\rm U}}
\newcommand{\SL}{{\rm SL}}

\newcommand{\be}{\begin{equation}}
\newcommand{\ee}{\end{equation}}

\title{%        %You can use \\ for explicit line-break
Boundary state of superstring in open string channel%
}

\author{%       %Use \scshape  for the family name
Yosuke \textsc{Imamura}$^{1,}$\footnote{E-mail: imamura@hep-th.phys.s.u-tokyo.ac.jp},  
Hiroshi  \textsc{Isono}$^{1,2,}$\footnote{E-mail: isono@hep-th.phys.s.u-tokyo.ac.jp~;~
after April 1, 2008: isono@phys.chuo-u.ac.jp}
and Yutaka \textsc{Matsuo}$^{1,}$\footnote{E-mail: matsuo@phys.s.u-tokyo.ac.jp}
}

\inst{%     %Affiliation, neglected when [addenda] or [errata]
$^1$Department of Physics, Faculty of Science, The University of Tokyo \\
Hongo 7-3-1, Bunkyo-ku, Tokyo 113-0033, Japan \\
$^2$Department of Physics, Chuo University \\
Kasuga 1-13-27, Bunkyo-ku, Tokyo 112-8551, Japan\footnote{
Address after April 1, 2008}
}

\abst{%         %this abstract is neglected when [addenda] or [errata]
We derive the boundary state of superstring in the open string channel.
The boundary state describes the superconformal field theory 
of open string emission and absorption by a D-brane.
We define the boundary state by conformal mappings from
the upper half plane with operators inserted at two points
corresponding to the corners of a semi-infinite strip.
We obtain explicit oscillator forms analytically for
the fermion and superconformal ghost sectors.
For the fermion sector we compare
the analytic result with the numerical result obtained
using the naive boundary condition.
}

\begin{document}

\maketitle

\section{Introduction} \label{intro.sec}
In two-dimensional conformal field theory, 
D-branes are described by boundary states.  
They belong to the closed string sector
and realize the boundary conditions associated with
the D-branes on the worldsheet. 
By taking the inner product with various closed string states,
one can describe the emission or absorption of
closed strings by the D-brane\cite{r:review}.

In previous papers \cite{IIM, IM}, we proposed an analog of the
boundary state in the open string channel\footnote{Similar states
are constructed in a different context in  Ref.~\citen{Others}.}.
As the original boundary state, the analog represents
the emission and absorption of the {\em open}
string by the D-brane (say $\Sigma$). Since the open string itself should be
attached to (other) D-brane(s) (say $\Xi_l$ and $\Xi_r$ for
the left and right ends of the open string, respectively), 
such a state is relevant
when these D-branes intersect,
$$
\Sigma\cap \Xi_{i}\neq \mathrm{null}\quad (i=l,r)\,.
$$
In the following we call such a state the open boundary state
or OBS in short.

While the massless part of a closed string boundary state
describes a gravitational $p$-brane solution,
the OBS describes a solitonic excitation of gauge fields
on the world volume of the D-brane $\Xi_{l,r}$.
In the previous notation, suppose we take $\Xi_l=\Xi_r\equiv \Xi$
as a D($p+4$)-brane and $\Sigma$ as a D$p$-brane embedded in
$\Xi$. Then we can extract the gauge field profile
of an instanton configuration on $\Xi$ from
the massless part of the associated OBS.

In previous papers \cite{IIM, IM}, 
we constructed the OBS for a bosonic
string and studied its properties in detail.
We derived the oscillator representations
for the bosonic field $X$ and the $(b,c)$-ghost.
The inner product between two OBSs represents an amplitude
whose worldsheet has a rectangular shape with its edges surrounded 
by various D-branes. We found that
the BRST invariance of the OBS imposes
nontrivial constraints on the D-branes that are attached to
the edges at each corner.

The purpose of this paper is to give a similar construction
of the OBS for the superstring case.   
This is a nontrivial step since there are
a few technical problems that do not arise in the bosonic case.
In \S\ref{boundary.sec}, we derive the boundary condition
that should be imposed on the OBS. It implies that the OBS
can be expressed in the form 
$\exp(\frac{1}{2} \sum_{rs} \psi_r K_{rs} \psi_s)|0\rangle$,
where $K_{rs}$ is an infinite-size matrix.
This approach, however, has an ambiguity
in the definition of the OBS because
some operator insertions
at the corners do not change the boundary conditions,
and the same boundary conditions may correspond to different
states.
In other words, the boundary conditions alone do not fix the 
matrix $K_{rs}$ uniquely.
In \S\ref{correlation.sect},
we solve this problem by 
another method developed in string field theory \cite{LPP}
where the correlation function is used to define the vertex.
Since the correlation function is uniquely defined once operators inserted
at the corners are given, one can uniquely fix the matrix $K$,
as shown in \S\ref{explicit.section}.
This also simplifies the derivation of the constraints from
the BRST invariance of the OBS through the CFT, 
as discussed in \S\ref{BRST.section}.
Finally in \S\ref{conclusion.section}, we provide
some applications of the OBS and point out some unsolved issues.

\section{Boundary conditions}\label{boundary.sec}
\begin{figure}[bt]
	\begin{center}
	\includegraphics[scale=0.7]{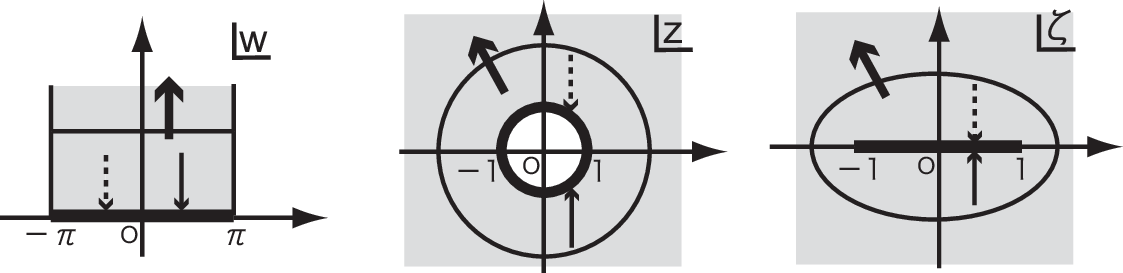}
	\end{center}
	\caption{Three sets of coordinates are used in the text. $w$
	is suitable for defining the boundary conditions of the OBS.
 	$\zeta=\cos w$ is used to define the correlation function of the
	boundary conformal field theory. 
	$z=e^{-iw}$ is used in the operator formalism.  
	The latter two are used in the next section.}
	\label{maps3.fig}
\end{figure}

The worldsheet diagram of an open string emitted from D-brane $\Sigma$
is given by a semi-infinite strip (see the first figure in Fig. {\ref{maps3.fig}}).
We use the variable $w(=\sigma+i\tau)$ to parametrize this region as
$0\leq\sig\leq\pi,~\tau\geq 0$.
The two endpoints of the open string
correspond to $\sigma=0$ and $\pi$, respectively, 
and are attached to D-branes
$\Xi_{l,r}$. D-brane $\Sigma$, from which the open string
is emitted, corresponds to the edge $0\leq\sig\leq\pi,~\tau=0$.
Let us call these boundaries
the left (or right) boundary and the bottom boundary, respectively.

The boundary conditions for the left and right edges are
\begin{eqnarray}
&{}&
\bar{\pd}X(\sig=0,\tau)=-\ep_l\pd X(\sig=0,\tau)\,,~~~~
\bar{\pd}X(\sig=\pi,\tau)=-\ep_r\pd X(\sig=\pi,\tau)\,, \label{blr} \\
&{}&
\tilde{\psi}(\sig=0,\tau)=i\eta_l\psi(\sig=0,\tau)\,,~~~~
\tilde{\psi}(\sig=\pi,\tau)=-i\eta_r\psi(\sig=\pi,\tau)\,, \label{flr}
\end{eqnarray}
and the boundary condition for the bottom edge is
\begin{eqnarray}
&{}&
\bar{\pd}X(\sig,\tau=0)=\ep_b\pd X(\sig,\tau=0)\,, \label{bb} \\
&{}&
\tilde{\psi}(\sig,\tau=0)=-\eta_b\psi(\sig,\tau=0)\,. \label{fb}
\end{eqnarray}
The parameters $\ep_{l,r,b}$ and $\eta_{l,r,b}$
take the values of $1$ or $-1$
and each sign describes Dirichlet or Neumann boundary conditions.
Note that the coefficient of the right-hand side of (\ref{flr})
has an extra factor of the imaginary unit $i$
if we compare it with (\ref{fb}).
This originates from the conformal transformation $w\rightarrow \pm iw$
for $\psi$ of conformal dimension 1/2.
Those factors disappear in the $\zeta$-plane, which will be discussed
in the next section (see (\ref{bc-zeta})), where boundary 
conditions are only set on the real axis.

We can replace antiholomorphic fields with holomorphic ones
using the doubling trick.
From the boundary conditions (\ref{blr}) and (\ref{flr}), we set
\begin{eqnarray}
&{}& \pd X(\sig,\tau)\equiv -\ep_l\bar{\pd}X(-\sig,\tau)
~~~~{\rm for~}-\pi<\sig<0\,, \label{bdb} \\
&{}& \psi(\sig,\tau)\equiv -i\eta_r\tilde{\psi}(-\sig,\tau)
~~~~{\rm for~}-\pi<\sig<0\,. \label{fdb}
\end{eqnarray}
We define the chiral fields on $-\pi \leq \sigma \leq \pi$.
$\pd X$ and $\psi$ must satisfy the following periodicity conditions
\begin{eqnarray}
\pd X(w+2\pi)=\ep_l\ep_r\pd X(w)\,,~~~~
\psi(w+2\pi)=-\eta_l\eta_r\psi(w)\,.
\end{eqnarray}
We extend the fields to the whole upper half plane using these conditions.
Combining the boundary conditions (\ref{bb}) and (\ref{fb})
with the doubling tricks (\ref{bdb}) and (\ref{fdb}),
we obtain the boundary conditions that define OBS $|B^o\rangle$,
\begin{eqnarray}
&{}& [\pd X(\sig,0)+\ep_l\ep_b\pd X(-\sig,0)]|B^o_X\rangle=0\,, \label{bobs} \\
&{}& [\psi(\sig,0)+i\eta_l\eta_b\ep(\sig)\psi(-\sig,0)]|B^o_{\psi}\rangle=0\,, \label{fobs-sign}
\end{eqnarray}
where $\ep(\sig)\equiv{\sign}(\sin{\sig})$.
The step function in (\ref{fobs-sign})
is the origin of the complication of the OBS for fermions.
It is indispensable to make the boundary condition
for $0<\sigma <\pi$ consistent with that
for $-\pi <\sigma <0$ with the pure imaginary factor.
Boundary conditions for the $(\beta,\gamma)$-superconformal ghost or
supercurrent $T_F$ take a similar form since the factor of $i$
originates from the half odd integer conformal weight of the chiral field.

\subsection*{\bf OBS for $X$ and $(b,c)$-ghost}
The boundary condition for $X$ (\ref{bobs}) was solved 
in Refs.~\citen{IIM} and \citen{IM}
and we obtained the explicit form of the OBS 
in terms of oscillators. Let us briefly recall the result.
Our notation for mode expansions is given in Appendix \ref{notation}.
The boundary conditions are
rewritten compactly as
\begin{eqnarray}
(\al_n+\ep_l\ep_b\al_{-n})|B^o_X\rangle^{\ep_b}_{\ep_l\ep_l}=0\,,
\label{boson-bc}
\end{eqnarray}
where index $n$ runs over positive integers when $\ep_l\ep_r=1$,
and over positive half odd integers when $\ep_l\ep_r=-1$.
The solution of this condition is
\begin{eqnarray} \label{boson-obs}
|B^o_X\rangle^{\epsilon_b}_{\epsilon_l\epsilon_r}
&\propto& \exp\left(
-\epsilon_l\epsilon_b\sum_{n>0} \frac{1}{2n}
\alpha_{-n}^2
\right)|\mbox{zero mode}\rangle_{\epsilon_l,\epsilon_r}^{\epsilon_b}\,,
\end{eqnarray}
where $|\mbox{zero mode}\rangle_{\epsilon_l,\epsilon_r}^{\epsilon_b}$ 
represents a product of the Fock vacuum times
and the zero-mode wave function.
When $(\epsilon_l,\epsilon_r)\neq (+1,+1)$,
we do not have a non-trivial zero mode and then
$|\mbox{zero mode}\rangle_{\epsilon_l,\epsilon_r}^{\epsilon_b}$
is simply the Fock vacuum.
For $(\epsilon_l,\epsilon_r)= (+1,+1)$,
the zero-mode wave function becomes nontrivial.
An appropriate choice is 
$\delta(p)$ for $\epsilon_b=1$ and
$\delta(x-x_0)$ with $x_0\in\mathbf{R}$ for $\epsilon_b=-1$.

OBS for the $(b,c)$-ghost system is 
defined by the following boundary conditions
\begin{eqnarray}
[c(\sig)+c(-\sig)]|B^o_{\rm gh}\rangle=
[b(\sig)-b(-\sig)]|B^o_{\rm gh}\rangle=0\,.
\end{eqnarray}
It is solved in terms of oscillators as \cite{IIM}
\begin{eqnarray} \label{bc-obs}
|B^o_{\rm gh}\rangle=\exp\left(\sum_{n>0}c_{-n}b_{-n}\right)c_0c_1
|\Omega\rangle\,,
\end{eqnarray}
where $|\Omega\rangle$ is the $\SL(2,\mathbf{R})$-invariant vacuum
for the $(b,c)$-ghost.

\subsection*{\bf OBS for $\psi$}
One may in principle apply the same strategy to
obtain the OBS for a fermion.
The boundary condition in terms of oscillators
is obtained by the Fourier transformation of the boundary condition (\ref{fobs-sign}),
\begin{eqnarray}\label{osc-bc1}
(\psi_r+\eta\sum_{-\infty<s<\infty}N_{rs}\psi_{s})|B^o_{\psi}\rangle=0\,,
\end{eqnarray}
where the index $r$ is $-\infty<r<\infty$,
$\eta=\eta_b\eta_l$,
and the infinite-dimensional matrix $N$ is
the Fourier transform of the step function.
In the NS sector, $N_{rs}$ takes the following form,
\begin{equation}
N_{rs}=\left\{
\begin{array}{l l}
0 \quad & (r+s=0)\\
\frac{1-(-1)^{r+s}}{\pi(r+s)} \quad& (r+s\neq 0)
\end{array}
\right.\,,
\end{equation}
where $r$ and $s$ run over half odd integers.
The matrix $N$ satisfies $\sum_s N_{rs}N_{st}=\delta_{r,t}$.
We decompose it into $2\times 2$ blocks,
\begin{eqnarray}
N
=\left(\begin{array}{@{}cc@{}}
N_{-r,-s} & N_{-r,s} \\
N_{r,-s} & N_{r,s}
\end{array}\right)
=\left(\begin{array}{@{}cc@{}}
n_{rs} & \tilde{n}_{rs} \\
-\tilde{n}_{rs} & -n_{rs}
\end{array}\right),
\end{eqnarray}
where indices $r$ and $s$ run over positive half odd integers.
$N^2=1$ implies that $n$ and $\tilde{n}$ satisfy 
\begin{eqnarray}\label{n-prop}
&{}&n^{2}-\tilde{n}^{2}=1,~~
n\tilde{n}=\tilde{n}n, ~~
n=n^{T},~~\tilde{n}=\tilde{n}^{T}.
\end{eqnarray}
We decompose this relation in terms of
the creation and annihilation parts,
\begin{eqnarray}\label{osc-bc2}
(\psi_{r}-\sum_{s>0}K_{rs}(\eta) \psi_{-s})|B^o_{\psi}\rangle=0\,,
\end{eqnarray}
where $r> 0$. The matrix $K$ is written in terms of $n$ and $\tilde n$ as
\begin{equation}\label{eq:K}
K(\eta)\equiv
\eta(1-\eta n)^{-1}\tilde{n}=-\eta\tilde{n}^{-1}(1+\eta n)
=-{K(\eta)}^{T}.
\end{equation}
The condition (\ref{osc-bc2}) is easily solved as
\begin{equation}
|B^o_{\psi}\rangle= \exp\left(\frac{1}{2}\sum_{r,s>0} K(\eta)
_{rs}\psi_{-r}\psi_{-s}\right)|\mathrm{vac}\rangle\,,
\end{equation}
where $|\mathrm{vac}\rangle$ is the Fock vacuum of NS sector.

The construction of the OBS for the Ramond sector is similar
although the treatment of the zero mode becomes tricky.
A related and even more serious issue is that it is possible
to insert some operators at the corners,
which do not affect the boundary condition.  To be more precise,
suppose we bosonize two Majorana fermions as (\ref{bosonize})
in Appendix \ref{notation}. 
Operators of the form $\exp\left(in H\right)$ ($n\in\mathbf{Z}$) 
are local with respect to the fermions and therefore
do not affect the boundary condition as long as
they are inserted at the corners.
It implies that the boundary condition alone does not fix the OBS uniquely
and we need the information of the correlation function.
In Appendix \ref{numerical.app}, 
the numerical treatment of $K(\eta)$ and its difficulty are explained.  
The problem is that the matrix elements of $K(1)$ 
obtained numerically are the same as 
those obtained analytically from (\ref{psiktilde})
and the formulae in Appendix \ref{formofk},
but in the case of $K(-1)$, they are different.

For this reason, we will not pursue this line of argument
in the following. Instead, we will use the technique of
string field theory, which solves these problems automatically.

\section{Definition of OBS through correlation functions}
\label{correlation.sect}
From this section, we will use the conformal field theory technique to derive the OBS
instead of using the boundary condition directly.

The relation between the two is similar to that between
the two alternative definitions of the interaction
vertex in string field theory. 
In the first definition, we express the gluing condition of the strings
by the delta functionals and derive the oscillator form of the vertex 
by solving the constraint. The treatment in the previous section
is analogous to this definition.
In the second definition, we
use the correlation functions \cite{LPP} of a single worldsheet
obtained by gluing strings using the vertex.
One can use a conformal transformation of this worldsheet
into a disk or the upper half plane and calculate the correlation
function by evaluating the disk amplitude.
The vertex (or more precisely the Neumann coefficient) 
is expressed in terms of the moments of this correlation function.

For the definition of the OBS, one can map the worldsheet of the semi-infinite strip
in the $w$-plane into the upper half plane 
with the insertions of the local fields
by (see Fig. \ref{maps3.fig})
\begin{eqnarray} \label{wtozeta}
\zeta=\cos w \,.
\end{eqnarray}
The three edges of the semi-infinite strip 
($\tau\geq 0, 0\leq \sigma\leq \pi$) are mapped
into three regions of the real axis in the $\zeta$-plane,
$\zeta>1, -1<\zeta<1,$ and $\zeta<-1$.

When $\phi$ is $\partial X$ ($h=1$)
and $\psi$ ($h=1/2$),
the boundary conditions 
(\ref{blr}), (\ref{flr}), (\ref{bb}), and (\ref{fb})
are replaced by 
\begin{eqnarray}\label{bc-zeta}
\bar{\pd}X(\bar{\zt})=\ep_i\pd X(\zt)\,,~~
\tilde{\psi}(\bar{\zt})=\eta_i\psi(\zt)\,,
\end{eqnarray}
where index $i$ represents $l, r, b$ for
$\zt>1$, $\zt<1$, and $-1<\zt<1$, respectively.
We use one of these conditions to replace the antichiral
field by the chiral field in the lower half plane
as the doubling trick.
Suppose we apply the doubling trick to the region $\zt>1$.  
The field $\partial X$ (or $\psi$)
will have a branch cut at $-1<\zt<1$ if the parameters satisfy
$\epsilon_l \epsilon_b=-1$ (or $\eta_l \eta_b=-1$).
This implies that we need an appropriate operator that changes the boundary
condition inserted at $\zeta=1$.  
Similarly, an operator insertion at $\zeta=-1$ is needed 
when $\epsilon_r \epsilon_b=-1$ (or $\eta_r \eta_b=-1$).
Let $O_{\pm 1}$ be the operators 
that are necessary at $\zt=\pm1$.
For the bosonic field $X^\mu$, the operator that changes the boundary condition
is the twist field $\sigma$ of conformal weight $1/16$,
which appears in the $\mathbf{Z}_2$ orbifold CFT \cite{DFMS}.  
For the fermion field $\psi^{\mu}$, 
the corresponding operator is the spin field $S_{\pm}$
(see Appendix \ref{notation} for notation).
$O_{\pm 1}$ is an appropriate product of 
the twist fields and the spin fields
which depends on the parameters $\epsilon_i$ and $\eta_i$.

We define the OBS for $\phi$ as the state
that reproduces the correlation function
\begin{eqnarray} \label{def}
&{}&
\langle{\Omega}|\phi_1^{(z)}(z_1)\cdots\phi_n^{(z)}(z_n)|B^o\rangle \nonumber\\
&{}&~~~~~~
=\langle\phi_1^{(\zeta)}(\zeta_1)\cdots\phi_n^{(\zeta)}(\zeta_n)O_{-1}O_{1}\rangle
\left(\frac{d\zeta_1}{dz_1}\right)^{h_1}\cdots
\left(\frac{d\zeta_n}{dz_n}\right)^{h_n}\,.
\end{eqnarray}
The left-hand side is the expression in the operator formalism,
whereas the right-hand side is the correlation function on the $\zeta$-plane
with operator insertions. 
The left-hand side can be computed using Wick's theorem with the propagator
define as the two-point function in the form (\ref{def}) (see Appendix \ref{proof.app}),
and the behaviors of the left- and right-hand sides at singularities are identical.
Since correlation functions are determined uniquely by the behavior at singularities,
this expression should be true for any number of insertions.
The coordinate $z$ is defined by $z=e^{-iw}$ (Fig. \ref{maps3.fig})
and is suitable for describing the CFT in the operator formalism.
The boundary associated with the OBS corresponds to a unit circle $|z|=1$.
Thus, the open string propagates from the unit circle to $z=\infty$. 
When $\phi$ is a free field such as $\partial X$ or $\psi$,
one can obtain the explicit form of the OBS from 
two-point correlation functions,
as will be explained in the next section.

There are some advantages of using the correlation functions
to define the OBS instead of the boundary condition
(\ref{osc-bc1}).
As we noted, the boundary condition at $0<\sigma<\pi$
does not fix the OBS uniquely since one may have many types of insertions 
at the corners ($\sigma=0,\pi$), which do not affect the boundary condition.
On the other hand, there is no ambiguity in the definition of the OBS (\ref{def})
since the correlation function is unique 
once we choose the operator insertions at the corners.
We can also avoid the technical difficulty in solving
equations including infinite-dimensional matrices
appearing in (\ref{osc-bc1}) and (\ref{osc-bc2}).
As we will see below,
we can easily obtain the explicit oscillator form of the OBS
by starting with (\ref{def}).

To describe the OBS for superstring,
we take the product
of the OBS for each field as
\begin{eqnarray}
|B^o_{\rm tot}\rangle=
\prod_{\mu=0}^9|B^o_{X^\mu} \rangle
\otimes\prod_{\mu=0}^9|B^o_{\psi^\mu}\rangle
\otimes|B^o_{\rm gh}\rangle |B^o_{\rm sgh}\rangle,
\end{eqnarray}
where
$|B^o_{X^\mu} \rangle$, $|B^o_{\psi^\mu}\rangle$, $|B^o_{\rm gh}\rangle$,
and $|B^o_{\rm sgh}\rangle$ are the OBS in the boson, fermion, 
$(b,c)$-ghost, and the $(\beta,\gamma)$-superconformal ghost sectors,
respectively.
The OBS in each sector is defined by (\ref{def}).

In superstring, the boundary conditions in
the bosonic and fermionic sectors
must be correlated to define the supercurrent $T_F$ consistently.
We introduce  $s_a=\pm1$ ($a=l,b,r$)  to represent the
boundary conditions for $T_F$,
\begin{equation}
\tilde T_F(\bar\zeta)=s_aT_F(\zeta),\quad
a=l,b,r,
\end{equation}
along the real axis of $\zeta$ as (\ref{bc-zeta}).
Since the supercurrent is given by $T_F=\psi^\mu\partial X_\mu$,
the relation $s_a=\epsilon_a^\mu \eta_a^\mu$ 
must hold for each pair of $\psi^\mu$ and $X^\mu$.
($\epsilon_a^\mu$ and $\eta_a^\mu$ are defined for each direction $\mu=0,\ldots,9$.)
We also need to choose an appropriate superconformal ghost sector 
for the vertices inserted at the corners
depending on the boundary conditions $s_a$.
If $s_a$ changes at $\zeta=\pm1$,
we need to insert a vertex operator
of the form $(\prod\sigma\prod S_{\pm})ce^{-\phi/2}$, 
which represents the R vacuum,
and otherwise we insert that of the NS vacuum of the form 
$(\prod 1\prod\sigma S_{\pm})ce^{-\phi}$.
We should carefully distinguish the sectors of the {\em vertex operators}
at the corners from the sector of the OBS itself.
The latter is defined by the combination of
the left and right boundary conditions,
while the sectors of the vertices are determined by the boundary conditions
at $\zeta=\pm 1$ (namely the corners in $w$-plane)
If the vertices are (NS,NS) or (R,R), OBS is in the NS sector,
and if the vertices are (NS,R) or (R,NS), the OBS is in the R sector.

\section{Explicit forms of OBS}
\label{explicit.section}
\subsection{Fermion sector}
In the following construction of the OBS in the fermion sector,
we need to use boundary changing operators for fermion fields.
For this reason it is convenient to define
complex fermions $\psi_\pm=(\psi_1\pm i\psi_2)/\sqrt2$ and bosonize them
as $\psi_\pm=e^{\pm iH}$.

Let $e^{ixH(-1)}$ and $e^{iyH(1)}$ be the inserted operators
at the two points $\zeta=\pm1$.
The two charges $x$ and $y$ should be chosen appropriately according to
the boundary conditions.
If the boundary condition changes at $\zeta=-1$,
$x$ must be a half odd integer, while if the boundary condition
does not change $x$ must be an integer.
The charge $y$ also should be chosen in the same way depending on
whether the boundary condition changes at $\zeta=+1$.

We can determine the OBS from the relation
\begin{eqnarray}
&&\langle\Omega|e^{-i(x+y)H(\infty)}\psi_+(z_1)\psi_-(z_2)|B^o_\psi\rangle_{xy}
\nonumber\\
&&=\langle e^{-i(x+y)H(\infty)}\psi_+(\zeta_1)\psi_-(\zeta_2)
e^{ixH(-1)}e^{iyH(1)}\rangle
\left(\frac{\partial\zeta_1}{\partial z_1}\right)^{1/2}
\left(\frac{\partial\zeta_2}{\partial z_2}\right)^{1/2}.
\label{corr-spin}
\end{eqnarray}
The insertion of $e^{-i(x+y)H}$ 
at infinity is necessary to cancel the total $U(1)$ charge.
The OBS $|B^o_\psi\rangle_{xy}$ 
satisfying this relation has a $U(1)$ charge $x+y$, and
we take the ansatz
\begin{equation}
|B_\psi^o\rangle_{xy}
=:\exp\left(\oint\frac{dz}{2\pi i}\oint\frac{dz'}{2\pi i}
\psi_-(z)K^{xy}(z,z')\psi_+(z')\right):e^{i(x+y)H(0)}|\Omega\rangle,
\label{bpsians}
\end{equation}
where $:\cdots:$ is the normal ordering defined on the `vacuum' state
$e^{i(x+y)H}|\Omega\rangle$.
Namely, we define creation and annihilation operators as
follows:
\begin{equation}
\mbox{creation: }\psi^+_{-x-y-r},\psi^-_{x+y-r}\,,\quad
\mbox{annihilation: }\psi^+_{-x-y+r},\psi^-_{x+y+r}\,,~~
\left(r=\frac{1}{2},\frac{3}{2},\ldots\right).
\label{creann}
\end{equation}
By substituting the ansatz (\ref{bpsians}) into the left-hand side of
(\ref{corr-spin}), we obtain
\begin{equation}
\mbox{l.h.s of (\ref{corr-spin})}
=D_{x+y}(z_1,z_2)
+\oint\frac{dz}{2\pi i}\oint\frac{dz'}{2\pi i}
D_{x+y}(z_1,z)K^{xy}(z,z')D_{x+y}(z',z_2),
\label{intint}
\end{equation}
where $D_{x+y}(z,z')$ is the propagator defined on the state $e^{i(x+y)H}|\Omega\rangle$,
and is given by
\begin{equation}
D_{x+y}(z,z')
=
\langle\Omega|e^{-i(x+y)H(\infty)}\psi_+(z)\psi_-(z')e^{i(x+y)H(0)}|\Omega\rangle
=\left(\frac{z}{z'}\right)^{x+y}\frac{1}{z-z'}.
\end{equation}
Let us assume that
the function $K^{xy}(z,z')$ is analytic in the region $|z|,|z'|>1$
and is damped sufficiently rapidly at infinity.
This will be confirmed
after we obtain an explicit form of the function $K^{xy}$.
On the basis of this assumption,
we can show that
the contour integrals in (\ref{intint}) pick up only the
contribution from the poles of the propagators at $z=z_1$ and $z'=z_2$,
and we obtain
\begin{equation}
\mbox{l.h.s of (\ref{corr-spin})}
=D_{x+y}(z_1,z_2)-K^{xy}(z_1,z_2).
\label{intint2}
\end{equation}
On the other hand, the right-hand side of (\ref{corr-spin}) is
easily computed as
\begin{equation}
\mbox{r.h.s of (\ref{corr-spin})}
=
D_{x+y}(z_1,z_2)
\frac{\sqrt{\left(1-\frac{1}{z_1^2}\right)\left(1-\frac{1}{z_2^2}\right)}}
{1-\frac{1}{z_1z_2}}
\left(\frac{1+\frac{1}{z_1}}{1+\frac{1}{z_2}}\right)^{2x}
\left(\frac{1-\frac{1}{z_1}}{1-\frac{1}{z_2}}\right)^{2y}.
\label{zetaplane}
\end{equation}
Comparing
(\ref{intint2}) and (\ref{zetaplane}), we obtain
the function $K(z_1,z_2)$ as
\begin{equation}
K^{xy}(z_1,z_2)=\frac{1}{z_1z_2}
\left(\frac{z_1}{z_2}\right)^{x+y}
{\cal K}^{xy}\left(\frac{1}{z_1},\frac{1}{z_2}\right),
\label{kfermi}
\end{equation}
where the function ${\cal K}^{xy}$ is defined by
\begin{equation}
{\cal K}^{xy}(u,v)=\frac{1}{u-v}
\left(
\frac{\sqrt{(1-u^2)(1-v^2)}}{1-uv}
\left(\frac{1+u}{1+v}\right)^{2x}
\left(\frac{1-u}{1-v}\right)^{2y}
-1
\right).
\label{psiktilde}
\end{equation}
${\cal K}^{xy}(u,v)$ is
analytic in the region $|u|,|v|<1$.
The potential singularity at $u=v$ is
canceled by the zero of the factor in the parentheses.
Using this fact we confirm the assumption
we used to perform the contour integrals in (\ref{intint}).
This behavior of ${\cal K}^{xy}$ also guarantees that
it can be expanded with respect to $u$ and $v$
in the region $|u|,|v|<1$
as
\begin{equation}
{\cal K}^{xy}(u,v)=\sum_{m,n=0}^\infty
K_{mn}^{xy}u^mv^n.
\label{kexp}
\end{equation}
A method of computing the explicit forms of the coefficients is
given in Appendix \ref{formofk}.
Using the coefficients $K_{mn}^{xy}$
we can explicitly give the OBS in the oscillator form.
\begin{equation}
|B^o_{\psi}\rangle_{xy}=\,:\exp\left(\sum_{n,m=0}^\infty 
\psi^-_{-m-1/2+x+y}K_{m,n}^{xy}\psi^+_{-n-1/2-x-y}\right):
e^{i(x+y)H(0)}|\Omega\rangle.
\end{equation}
Note that the indices of the fermion oscillators run over all the creation operators
defined in (\ref{creann}).

The matrix $K^{xy}$ thus defined should agree with $K^{xy}$ in (\ref{eq:K})
expressed in terms of infinite-dimensional matrices. 
We numerically confirmed this in Appendix \ref{numerical.app}.
Note that there exist discrepancies when $(x,y)\neq(0,0)$,
which are due to the nontrivial operator insertion at the corners.

\subsection{Superconformal ghost sector}
Let us construct the OBS in the superconformal ghost sector.
We denote the OBS defined using
the inserted operators $e^{p\phi(-1)}$ and $e^{q\phi(1)}$
by $|B^o_{\rm sgh}\rangle_{pq}$,
where $p$ and $q$ are the pictures of the inserted vertex operators.
The picture of the OBS itself is $p+q$.
The OBS in the superconformal ghost sector can be determined
using the following relation:
\begin{eqnarray}
&&
\langle\Omega|e^{-(p+q+2)\phi(\infty)}\gamma(z_1)\beta(z_2)|B^o_{\rm sgh}\rangle_{pq}
\nonumber\\&&
=\langle e^{-(p+q+2)\phi(\infty)}\gamma(\zeta_1)\beta(\zeta_2)
e^{p\phi(-1)}e^{q\phi(1)}\rangle
\left(\frac{d\zeta_1}{dz_1}\right)^{-1/2}
\left(\frac{d\zeta_2}{dz_2}\right)^{3/2}.
\label{scsectordef}
\end{eqnarray}
We take the ansatz
\begin{equation}
|B^o_{\rm sgh}\rangle_{pq}
=\exp\left(\oint\frac{dz}{2\pi i}\oint\frac{dz'}{2\pi i}
\beta(z)\tilde K^{pq}(z,z')\gamma(z')\right)
e^{(p+q)\phi(0)}|\Omega\rangle.
\end{equation}
By substituting this ansatz into the left-hand side of (\ref{scsectordef}),
we obtain
\begin{equation}
\mbox{l.h.s of (\ref{scsectordef})}
=D_{p+q}(z_1,z_2)-\tilde K^{pq}(z_1,z_2),
\label{scglhs}
\end{equation}
where $D_{p+q}$ is the propagator defined by
\begin{equation}
D_{p+q}(z,z')=
\langle\Omega|e^{-(p+q+2)\phi(\infty)}\gamma(z)\beta(z')e^{(p+q)\phi(0)}|\Omega\rangle
=\left(\frac{z}{z'}\right)^{p+q}\frac{1}{z-z'}.
\end{equation}
The right-hand side of (\ref{scsectordef}) is easily computed as
\begin{equation}
\mbox{r.h.s of (\ref{scsectordef})}
=D_{p+q}(z_1,z_2)
\frac{\sqrt{\left(1-\frac{1}{z_1^2}\right)\left(1-\frac{1}{z_2^2}\right)}}{1-\frac{1}{z_1z_2}}
\left(\frac{1+\frac{1}{z_1}}{1+\frac{1}{z_2}}\right)^{2p-1}
\left(\frac{1-\frac{1}{z_1}}{1-\frac{1}{z_2}}\right)^{2q-1}.
\label{scgrhs}
\end{equation}
By comparing (\ref{scglhs}) and (\ref{scgrhs})
we obtain
\begin{equation}
\tilde K^{pq}(z_1,z_2)=\frac{1}{z_1z_2}\left(\frac{z_1}{z_2}\right)^{p+q}
{\cal K}^{p-1/2,q-1/2}\left(\frac{1}{z_1},\frac{1}{z_2}\right),
\end{equation}
where ${\cal K}$ is the function defined in (\ref{psiktilde}).
Using the expansion coefficients in (\ref{kexp})
we can obtain the oscillator form of the OBS:
\begin{equation}
|B^o_{\rm sgh}\rangle_{pq}=\exp\left(
\sum_{m,n=0}^\infty\beta_{-m+p+q-3/2}K_{mn}^{p-1/2,q-1/2}\gamma_{-n-p-q+1/2}
\right)
e^{(p+q)\phi(0)}|\Omega\rangle.
\end{equation}

\section{BRST invariance of OBS}
\label{BRST.section}
\begin{figure}[bt]
	\begin{center}
	\includegraphics[scale=1.0]{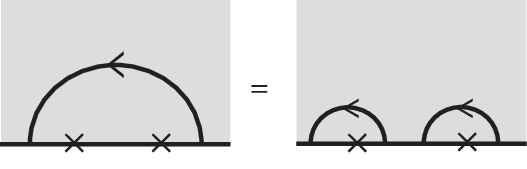}
	\end{center}
	\caption{Integration contour of the BRST current and its deformation.}
	\label{charge.fig}
\end{figure}

In this section, we derive constraints
from the BRST invariance of the  OBS.
Let $Q_B$ be the BRST charge and $j_B$ 
be the corresponding BRST current.
In the bosonic case \cite{IIM}, we have seen that the BRST
invariance
\begin{equation}\label{e:BRST}
Q_B|B^o_{\rm tot}\rangle=0
\end{equation}
implies that the
number of twist fields (namely the number of ND sectors)
at each corner must be 16. The computation in the operator
formalism performed in Ref.~\citen{IIM} was complicated 
but can be understood more directly.
From the correspondence between the operator formalism
and the correlation function (\ref{def}), 
the insertion of $Q_B=\int dz j_B(z)$
in front of $|B^o_{\rm tot}\rangle$ is equivalent to the insertion of 
$\int d\zeta j_B(\zeta)$, where the contour surrounds
two points $\zeta=\pm 1$
associated with two corners (see Fig. \ref{charge.fig}).
As shown in the figure, this contour can be deformed to two
semicircles around $\zt=\pm 1$.  
The BRST invariance (\ref{e:BRST}) is thus reduced to the BRST
invariance of the operators inserted there.
The BRST invariance requires that the dimension of the insertion is zero.

Here we restrict ourselves to the boundary changing operators
of the form, $c(\pm 1) 
\prod_{\mu\in \mathrm{DN}} \sigma^\mu(\pm 1)$,
where $\mu$ runs over the Dirichlet-Neumann directions.
Since the conformal dimension of $\sigma$ is $1/16$ and that of
$c$ is $-1$, the number of twist fields should be $16$.
Note that the Dirichlet-Neumann sector here is that for the
open string which interpolates between the bottom
and the left (or right) boundaries. For example, suppose
the D-branes ($\Xi_{l,r}$) where the open string is attached is D25-brane,
the D-brane ($\Sigma$) described by the OBS should be the D9-brane.

We can apply a similar method to the superstring case. 
We have to be careful in the fact that the open string
interpolates the bottom and left (or right) boundaries and
we have to specify the NS or R sectors for such open strings.
We will call these sectors the $\mathrm{NS}^c$ and $\mathrm{R}^c$ sectors,
where the superscript $c$ implies the corner.

For the $\mathrm{NS}^c$ sector, the natural ghost insertion is
$ce^{-\phi}$ of conformal dimension $-1/2$.
On the other hand, in the DN direction we must
insert $\sig S$ of conformal dimension $1/8$.
In the DD and NN sectors, we insert no operators in the matter
sector.  Therefore to cancel conformal dimensions, 
the number of DN directions must be four.

For the $\mathrm{R}^c$ sector, 
the ghost insertion is $ce^{-\phi/2}$ of
conformal dimension $-5/8$. In the matter sector,
in the DN direction we insert $\sigma$
and in the DD and NN directions we insert $S$.
In both cases, the operator from the matter sector has dimension $1/16$.
Therefore, the conformal dimension always cancels between the matter sector
and the ghost sector.  Namely, there is no constraint
originating from the R sector.

To summarize, if we require the BRST invariance of the OBS 
in both the NS and R sectors, 
the number of DN directions should be four.  
This coincides with the condition of the intersecting D-branes, 
where the open strings 
that intertwine the two D-branes
have massless modes with no momenta
and the spacetime supersymmetry is partially preserved.
Note that this strong result originates from 
the restriction on the boundary changing operators 
with the twist and spin fields.
If we use more general operators, 
we can construct an OBS for more general D-brane configurations.

\section{Conclusion and discussion}
\label{conclusion.section}
In this study, we carried out the explicit construction of OBS
for superstring.  We have encountered a few technical challenges
compared with the bosonic case, \cite{IIM, IM}
which include the ambiguity of operator insertion
at the corners.  Nevertheless, we obtained exact expressions
for both the fermion and the superconformal ghost.  The concrete form
of OBS is more complicated than that for the bosonic case.
The computation of the inner product between OBSs,
which was easy in the bosonic case, becomes technically
more difficult and we could not carry it out in this study.

There are a few applications of the OBS 
that may be interesting in the future.  
In \S\ref{intro.sec}, 
the relation between our OBS and the instanton profile 
in the D$(p+4)$-D$p$-brane system
was briefly indicated.
For a closed string,
it is well-known \cite{DiVecchia:1997pr} that 
the long-distance behavior of
the classical supergravity solutions of D-branes can be constructed
from the massless part of the corresponding boundary states.
It may be thought that, just as in the closed string case,
we can reproduce the soliton profile for source D-branes 
in higher-dimensional D-branes 
by extracting the massless part in the OBS. 
Note that such an analysis
has been carried out by Billo et al.\cite{0211250} for D$3$-D($-1$) systems, 
although the concept of the OBS was not introduced in that study. 
They computed 
a disk amplitude with mixed boundary conditions
$\langle V_{\mu}\cal{O}_{+}\cal{O}_{-}\rangle$,
where $V_{\mu}$ is the vertex operator of the
gauge field on the D$3$-brane, 
and $\cal{O}_{+}$ and $\cal{O}_{-}$ are vertex operators 
that correspond to
the massless scalar fields in  3-$(-1)$ and $(-1)$-3 open strings, respectively,
and these vertex operators are inserted on the boundary of the disk.
Since they do not  have momentum, the vertex operators coincide with
the operators $\mathcal{O}_\pm$ in the definition of OBS.
This disk amplitude is in the form of (\ref{def})
with $n=1$ and is equivalent to 
$A_\mu=\langle \Omega| V_{\mu}\frac{1}{L_0}|B^o\rangle$.
Here $\langle \Omega| V_{\mu}$
is the massless state of open strings on the D3-branes
and $|B^o\rangle$ is the OBS for D($-1$)-branes.
The authors of Ref.~\citen{0211250} 
showed that the correlator  $\langle V_{\mu}\cal{O}_{+}\cal{O}_{-}\rangle$
reproduces the instanton profile on the D3-branes. This implies that
the same statement for $A_\mu$ can be written in terms of the OBS.
In this way, the OBS can be regarded as a generalization of the
concept of instanton configuration, that contains the information
of all the open string excitation.

Another future direction is to explore the relation with string
field theory (SFT).  So far, the boundary state has been mainly used in
SFT as the source term \cite{HH}. Since the usual boundary state 
belongs to the closed string sector,
we need closed SFT to 
introduce such coupling. However, our understanding
of closed SFT is not complete.
On the other hand, the OBS can be used as
the source term for open SFT, where we have a standard
formulation.  Namely, Witten's action is redefined in the presence of
a D-brane as
\begin{eqnarray}
S=\frac{1}{2}\int \Psi \star Q\Psi + \frac{g}{3}\int \Psi\star\Psi\star \Psi
 + \int \Psi\star |B^o\rangle\,.
\end{eqnarray}
This gives a natural introduction of the D-brane in open SFT.
It will be interesting to explore the consequences of such coupling.
For example, the idempotency relation,
\begin{eqnarray}
|B^o\rangle\star|B^o\rangle \propto |B^o\rangle\,,
\end{eqnarray}
which was shown in Ref.~\citen{IM} (as a generalization of the closed string relation
\cite{KMW}) appears as a consistency condition of such coupling.
We will return to this issue in a forthcoming paper \cite{IIM3}.

\section*{Acknowledgements}
We would like to thank  K. Murakami, I. Kishimoto, T. Takahashi, and S. Teraguchi
for their interesting comments.
The authors thank the Yukawa Institute for Theoretical Physics at Kyoto University. Discussions during the YITP workshop YITP-W-07-05 on ``String Theory and Quantum Field Theory'' were useful in completing this work.
Y.M. is partially supported by
a Grant-in-Aid for Scientific Research (C) (\#16540232).
Y.I. is partially supported by
a Grant-in-Aid for Young Scientists (B) (\#19740122) from the Japan
Ministry of Education, Culture, Sports,
Science and Technology.
H.I. is supported in part by a JSPS Research Fellowship
for Young Scientists ($11\cdot 11762$).

\appendix
\section{Notation} \label{notation}
Here we give the notation used in this paper.
In the bosonic string sector,
the mode expansions of $\pd X$ with various boundary conditions
are given by
\begin{eqnarray}
&&X^{(NN)}(w,\bar w)=\hat x-\alpha' \hat p(w-\bar w)
+i\left(\frac{\alpha'}{2}\right)^{\frac{1}{2}} \sum_{m\neq 0}\frac{1}{m}
\alpha_m(e^{imw}+e^{-im\bar w})\,,\\
&&X^{(DD)}(w,\bar w) =
x+\frac{y-x}{2\pi}(w+\bar w)+
i\left(\frac{\alpha'}{2}\right)^{\frac{1}{2}} \sum_{m\neq 0}\frac{1}{m}
\alpha_m(e^{imw}-e^{-im\bar w})\,,\\
&& X^{(DN)}(w,\bar w) =x +i\left(\frac{\alpha'}{2}\right)^{\frac{1}{2}}
\sum_{r\in Z+1/2}\frac{1}{r}\alpha_r(e^{irw}-e^{-ir\bar w})\,,\\
&& X^{(ND)}(w,\bar w) =x +i\left(\frac{\alpha'}{2}\right)^{\frac{1}{2}}
\sum_{r\in Z+1/2}\frac{1}{r}\alpha_r(e^{irw}+e^{-ir\bar w})\,.
\end{eqnarray}
The commutation relations for mode variables are
\begin{eqnarray}
 [\alpha_n,\alpha_m]=n\delta_{n+m,0}\,,\quad
 [\hat x,\hat p]=i\,.
\end{eqnarray}

Let us review the notation of the fermionic string sector.
A fermionic string is represented by a worldsheet field $\psi^{\mu}$,
where $\mu$ is the spacetime index.
(We often omit this spacetime index.)
The mode expansions and the operator product expansion are
\begin{eqnarray}
&{}&
\psi(z)=\sum_{r}\frac{\psi_r}{z^{r+1/2}}\,,~~~~
\psi(w)=(-i)^{1/2}\sum_{r}\psi_re^{irw}\,, \\
&{}&
\psi(z_1)\psi(z_2)\sim\frac{1}{z_1-z_2}\,,~~~~
\{\psi_r,\psi_s\}=\delta_{r+s,0}\,,
\end{eqnarray}
where the index $r$ is an integer or half odd integer, and
is determined by the periodicity condition of $\psi$.
When considering the spin operator,
one spin field involves two spacetime directions,
namely the Dirac fermions on the worldsheet are necessary,
\begin{eqnarray}
\psi_{\pm}\equiv\frac{1}{\sqrt{2}}(\psi^1\pm i\psi^2)\,,~~~~
\psi_+(z_1)\psi_-(z_2)\sim\frac{1}{z_1-z_2}\,,~~~~
\{\psi^+_r,\psi^-_s\}=\delta_{r+s,0}\,,
\end{eqnarray}
where indices $\pm$ represent the $\U(1)$ charge $\pm 1$ of the fields,
defined by the current $j_{\U(1)}=\psi_+\psi_-$.

Their bosonized forms are defined as,
\begin{eqnarray}
\psi_{\pm}\cong e^{\pm iH}\,,~~~~
H(z_1)H(z_2)\sim -\ln(z_1-z_2)\,.
\end{eqnarray}
The spin fields are defined as,
\begin{eqnarray}\label{bosonize}
S_{\pm}\cong\exp\left(\pm\frac{i}{2}H\right).
\end{eqnarray}
The background charge is equal to 0.
Thus, the net $\U(1)$ charge of all the fields
between $\langle\Omega|$ and $|\Omega\rangle$ in the correlation function
must be equal to $0$.
Let $|x\rangle$ be $e^{ixH(0)}|\Omega\rangle$,
which is used in the construction of the OBS.
Then the dual state $\langle\wt{x}|$ is $\langle\Omega|e^{-ixH(\infty)}$.

Our notations used for the $(\bt,\ga)$-superconformal ghost system 
is the standard notation \cite{FMS}.
The conformal weight of $\ga$ ($\bt$) is $-1/2$ ($3/2$).
The background charge of this system is $Q=2$.
Their mode expansions and the operator product expansions are
\begin{eqnarray}
&{}&
\gamma(z)=\sum_{r}\frac{\gamma_r}{z^{r-1/2}}\,,~~~~
\beta(z)=\sum_{r}\frac{\beta_r}{z^{r+3/2}}\,, \\
&{}&
\gamma(w)=(-i)^{-1/2}\sum_{r}\gamma_re^{irw}\,, ~~~~
\beta(w)=(-i)^{3/2}\sum_{r}\beta_re^{irw}\,, \\
&{}&
\gamma(z_1)\beta(z_2)\sim\frac{1}{z_1-z_2}\,,~~~~
\beta(z_1)\gamma(z_2)\sim -\frac{1}{z_1-z_2}\,,~~~~
[\gamma_r,\beta_s]=\delta_{r+s,0}\,.
\end{eqnarray}
The bosonizations are
\begin{eqnarray}
&{}&
\gamma\cong e^{\phi}\eta\,,~~~~\beta\cong (\pd\xi)e^{-\phi}\,, \\
&{}&
\phi(z_1)\phi(z_2)\sim -\ln(z_1-z_2)\,,~~~~\xi(z_1)\eta(z_2)\sim\frac{1}{z_1-z_2}\,.
\end{eqnarray}
In our calculation, $(\eta,\xi)$ does not appear.
The $\U(1)$ charge is defined by the $\U(1)$ current $j=-\bt\ga=-\pd\phi$.
Thus, $\ga$ ($\bt$) has $\U(1)$ charge $1$ ($-1$).
In the bosonized form, we can define the following vacua
\begin{eqnarray}
|q\rangle\equiv e^{q\phi(0)}|\Omega\rangle\,.
\end{eqnarray}
By definition, $|q\rangle$ has $\U(1)$ charge $q$.
The conformal weight of $e^{q\phi}$ is $-q(q+2)/2$.
The background charge of this system is $Q=2$.
Thus the net $\U(1)$ charge of all the fields
between $\langle\Omega|$ and $|\Omega\rangle$ in the correlation function
must be equal to $-2$.
Then, the dual bra vacuum $\langle\wt{q}|$ satisfying $\langle\wt{q}|q\rangle=1$
is defined by $\langle\wt{q}|\equiv\langle\Omega|e^{(-2-q)\phi(\infty)}$.

\section{OBS in the bosonic string sector} \label{obs-boson}
The simplest way to obtain the OBS in the bosonic sector
is to directly use the boundary condition (\ref{boson-bc}).
In Ref.~\citen{IIM}, the OBS is obtained in this way.
It is, of course, possible to use the
method of conformal mapping to construct the OBS, as
shown in \S\ref{explicit.section} 
for the fermion and superconformal ghost sectors.
We demonstrate the construction in the following.

To make the derivation as similar as possible to the
other cases,
we here use a complex chiral boson $Z$ and its conjugate $\bar Z$ with the OPE
\begin{equation}
Z(z)\bar Z(z')\sim -\log(z-z')\,.
\end{equation}
The inserted operators used in the definition of the OBS
in this case are $\sigma^{2p}(-1)$ and $\sigma^{2q}(1)$,
where $\sigma(z)$ is the twist operator for the boson fields $Z$ and $\bar Z$.
The numbers $p,q=0,1/2$ are chosen according to the boundary conditions.
We use the notation
\begin{equation}
\sigma^1\equiv \sigma,\quad
\sigma^0\equiv\mbox{identity operator}\,.
\end{equation}
It is also convenient to define
\begin{equation}
\sigma^2(z)\equiv\delta(Z(z))\sim 
\lim_{\epsilon\rightarrow 0}\sigma(z+\epsilon)\sigma(z)\,.
\end{equation}
This operator changes the $\SL(2,\mathbf{R})$ vacuum into the position eigenstate
\begin{equation}
\sigma^2(0)|\Omega\rangle=|Z=0\rangle\,.
\end{equation}
Using this notation, the defining equation of the OBS is given as
\begin{eqnarray}
&&
\langle\Omega|\sigma^{2-2p-2q}(\infty)Z(z_1)\partial\bar Z(z_2)|B^o_Z\rangle_{pq}
\nonumber\\
&&=\langle\sigma^{2-2p-2q}(\infty)Z(\zeta_1)\partial\bar Z(\zeta_2)
\sigma^{2p}(-1)\sigma^{2q}(1)\rangle
\frac{\partial\zeta_2}{\partial z_2}\,.
\label{defbz}
\end{eqnarray}
We insert $\sigma^2$ at infinity when $p=q=1/2$.
Although this is not necessary to obtain a nonvanishing amplitude,
it is convenient because
it removes the divergence associated with the infinite volume of the $Z$ space,
and it makes expression (\ref{defbz}) similar to the corresponding equations
in the fermion and superconformal ghost cases.
We take the following ansatz:
\begin{equation}
|B^o_Z\rangle_{pq}
=:\exp\left(\oint\frac{dz}{2\pi i}\oint\frac{dz'}{2\pi i}
\partial\bar Z(z)K(z,z')Z(z')
\right):\sigma^{2p+2q}(0)|\Omega\rangle\,.
\end{equation}
By substituting this ansatz into the left-hand side of
(\ref{defbz}) we obtain
\begin{equation}
\mbox{l.h.s. of (\ref{defbz})}
=D_{p+q}(z_1,z_2)-K(z_1,z_2)\,,
\label{bosonlhs}
\end{equation}
where the propagator $D_{p+q}$ is
given by
\begin{equation}
D_{p+q}(z,z')
\equiv\langle\Omega|\sigma^{2-2p-2q}(\infty)Z(z)\partial\bar Z(z')\sigma^{2p+2q}(0)|\Omega\rangle
=\left(\frac{z}{z'}\right)^{p+q}\frac{1}{z-z'}\,.
\end{equation}
The right-hand side of (\ref{defbz}) is
\begin{equation}
\mbox{r.h.s. of (\ref{defbz})}
=D_{p+q}(z_1,z_2)\frac{\sqrt{(1-\frac{1}{z_1^2})(1-\frac{1}{z_2^2})}}{1-\frac{1}{z_1z_2}}
\left(\frac{1+\frac{1}{z_1}}{1+\frac{1}{z_2}}\right)^{2p-\frac{1}{2}}
\left(\frac{1-\frac{1}{z_1}}{1-\frac{1}{z_2}}\right)^{2q-\frac{1}{2}}\,.
\label{rhsboson}
\end{equation}
This is similar to (\ref{zetaplane}) and (\ref{scgrhs}),
and differs from them only by the powers of the last two factors,
which are due to the difference in the conformal dimensions of the fields.
By comparing (\ref{bosonlhs}) and (\ref{rhsboson}),
we obtain
\begin{equation}
K(z_1,z_2)=\frac{1}{z_1z_2}\left(\frac{z_1}{z_2}\right)^{p+q}
{\cal K}^{p-1/4,q-1/4}\left(\frac{1}{z_1},\frac{1}{z_2}\right)\,.
\end{equation}
In this case, the function
${\cal K}^{p-1/4,q-1/4}$ does not include square roots and
is simplified to
\begin{equation}
{\cal K}^{p-1/4,q-1/4}(u,v)=
\frac{(-)^{2q}u^{(1-2p)(1-2q)}v^{4pq}}{1-uv}\,.
\end{equation}
We can easily compute the expansion coefficients of this function
and obtain the explicit form of OBS in the bosonic sector
\begin{equation}
|B^o_Z\rangle_{pq}=
\exp\left(-(-)^{2q}\sum_{r>0}\frac{\bar\alpha_{-r}\alpha_{-r}}{r}\right)
\sigma^{2p+2q}(0)|\Omega\rangle\,,
\end{equation}
where the index $r$ runs over positive integers (positive half odd integers)
when $2(p+q)$ is even (odd).

%%%%%%%%%%%%%%%%%%%%%%%%%%%%%%%%%%%%%%%%%%%%%%%%%%%%%%
\section{$n$-point functions} \label{proof.app}
When we define the OBS, we used only the $2$-point function
\begin{equation}
{\cal D}(z_1,z_2)=\langle 0|\phi(z_1)\phi(z_2)|B^o\rangle\,,
\label{cald}
\end{equation}
where $\phi$ are various types of fields, $|B^o\rangle$
is the OBS in the corresponding sector, and $\langle0|$ is an
suitable vacuum state.
To compute general $n$-point correlation functions
of the form
\begin{equation}
\langle 0|\phi_1(z_1)\phi_2(z_2)\cdots\phi_n(z_n)|B^o\rangle\,,
\label{npoint}
\end{equation}
we can use Wick's theorem with the propagator (\ref{cald}).
Namely, the amplitude is given as the sum of the contributions of
all pairings of the operators $\phi_k$ ($k=1,\ldots,n$),
and the contribution of each pairing is
obtained by replacing each pair by the propagator
(\ref{cald}).
We prove this fact.
The proof applies to any free field
if we replace $\phi$, $|B^o\rangle$, $\langle 0|$, etc.,
by appropriate fields and states.
We will not distinguish them here.

To prove the above statement,
the following identity is useful:
\begin{eqnarray} \label{prop-z}
\phi(z)|B^o\rangle
=\oint_{|z'|>|z|}\frac{dz^{\prime}}{2\pi i}
\phi^{\rm cr}(z^{\prime})|B^o\rangle
{\cal D}(z',z)\,,
\end{eqnarray}
where the integration contour is a circle of radius $|z^{\prime}|>|z|$
and $\phi^{\rm cr}$ is the creation part of $\phi$
on the vacuum $|0\rangle$, which is the vacuum state used in
the ansatz $|B^o\rangle=:\exp(\phi K\phi):|0\rangle$.
Let us first prove this equation.
We decompose the operator $\phi(z)$ on the left-hand side to
the annihilation part $\phi^{\rm an}(z)$ and the creation part $\phi^{\rm cr}(z)$.
For the annihilation part,
using $|B^o\rangle=:\exp(\phi K\phi):|0\rangle$,
we obtain
\begin{eqnarray}\label{c4}
\phi^{\rm an}(z)|B^o\rangle
&=&
\oint_{|z'|<|z|}\frac{dz'}{2\pi i}\oint_{|z''|<|z|}\frac{dz''}{2\pi i}\phi^{\rm cr}(z')
|B^o\rangle K(z',z'')D(z'',z)
\nonumber\\
&=&
-\oint_{|z'|>|z|}\frac{dz'}{2\pi i}\phi^{\rm cr}(z')|B^o\rangle K(z',z)\,,
\label{annpart}
\end{eqnarray}
where $D(z,z')$ is the propagator defined by 
$D(z,z')=\langle 0|\phi(z)\phi(z')|0\rangle$.
In (\ref{c4}), we performed $z''$ integral in the same way as for the integral
in (\ref{intint}).
We deformed the contour outward
and used the fact that the integral around $z''=\infty$ vanishes.
Only the pole of the propagator $D(z'',z)$ contributes to this integral.
For the $z'$ integral, we deformed the contour
from a circle inside $z$ to a circle outside $z$ by using
the regularity of the function $K(z',z)$ at $z'=z$.

The creation operator part is rewritten as
\begin{eqnarray}
\phi^{\rm cr}(z)|B^o\rangle
=\oint_{|z'|>|z|}\frac{dz'}{2\pi i}\phi^{\rm cr}(z')|B^o\rangle D(z',z)\,,
\label{crepart}
\end{eqnarray}
where we used the operator identity 
$\oint_{|z'|>|z|} \phi^{\rm cr}(z')D(z',z)=\phi^{\rm cr}(z)$.
If we use the relation like (\ref{intint2})
we can see that the sum of (\ref{annpart}) and (\ref{crepart})
is the right-hand side of (\ref{prop-z}),
and we have proven relation (\ref{prop-z}).

We apply formula (\ref{prop-z}) to the rightmost operator
in the correlation function (\ref{npoint}) and obtain
\begin{equation}
\oint_{|z_{n-1}|>|z|>|z_n|}\frac{dz}{2\pi i}
\langle 0|\phi_1(z_1)\phi_2(z_2)\cdots\phi_{n-1}(z_{n-1})\phi_n^{\rm cr}(z)|B^o\rangle
{\cal D}(z,z_n)\,.
\end{equation}
By the Wick contraction of the operator $\phi_n^{\rm cr}(z)$ and
other operators $\phi_k(z_k)$ ($k=1,\ldots,n-1$),
we obtain
\begin{equation}
\oint_{|z_{n-1}|>|z|>|z_n|}\frac{dz}{2\pi i}
\sum_{k=1}^{n-1}\pm\langle 0|\phi_1(z_1)\cdots\check\phi_k(z_k)\cdots\phi_{n-1}(z_{n-1})
|B^o\rangle D(z_k,z){\cal D}(z,z_n)\,.
\end{equation}
The sign in the summand should be chosen appropriately
according to the statistics of the operators.
By deforming the integration contour,
the summand can be rewritten as the sum of the pole contributions,
\begin{equation}
\sum_{k=1}^{n-1}\pm\langle 0|\phi_1(z_1)\cdots\check\phi_k(z_k)\cdots\phi_{n-1}(z_{n-1})|B^o\rangle
{\cal D}(z_k,z_n)\,.
\end{equation}
If we iterate this procedure we obtain the correlation function
as a combination of propagators (\ref{cald}).

%%%%%%%%%%%%%%%%%%%%%%%%%%%%%%%%%%%%%%%%%%%%%%%%%%%%%%
\section{Explicit form of ${\cal K}$}\label{formofk}

In this appendix we briefly explain how we expand the function
\begin{equation}
{\cal K}^{xy}(u,v)
=\frac{1}{u-v}
\left(
\frac{\sqrt{(1-u^2)(1-v^2)}}{1-uv}
\left(\frac{1+u}{1+v}\right)^{2x}
\left(\frac{1-u}{1-v}\right)^{2y}
-1
\right).
\end{equation}
We first divide ${\cal K}^{xy}$ into two parts
${\cal K}^1$ and ${\cal K}^2$ defined by
\begin{eqnarray}
{\cal K}^{xy}={\cal K}^1+{\cal K}^2,\quad
{\cal K}^1\equiv\frac{P(u)Q(v)-1}{u-v},\quad
{\cal K}^2\equiv-\frac{P(u)Q(v)}{1-uv}\,.
\end{eqnarray}
$P(u)$ and $Q(v)$ are functions of $u$ and $v$, respectively:
\begin{equation}
P(u)=(1+u)^{2x+1/2}(1-u)^{2y-1/2},\quad
Q(v)=(1+v)^{-2x-1/2}(1-v)^{-2y+1/2}\,.
\end{equation}
If the functions
${\cal K}^1$ and ${\cal K}^2$ did not include the factors
$1/(v-u)$ and $1/(1-uv)$, which are not factorized into
functions of $u$ and $v$,
we would easily obtain the expansions of ${\cal K}^1$ and ${\cal K}^2$.
The unwanted factors can be removed by applying appropriate
differential operators to these functions (we follow a similar computation 
in Ref.~\citen{GJ3}).
\begin{eqnarray}
(u\partial_u+v\partial_v+1){\cal K}^1
&=&
\left[
  \frac{1/2+2x}{(1+u)(1+v)}
 +\frac{1/2-2y}{(1-u)(1-v)}
 \right]P(u)Q(v),\\
(u\partial_u-v\partial_v-2x-2y){\cal K}^2
&=&
 \left[
  \frac{1/2+2x}{(1+u)(1+v)}
 -\frac{1/2-2y}{(1-u)(1-v)}
 \right]P(u)Q(v).
\end{eqnarray}
The right-hand side of these equations
consists of only factorized terms,
and from these equations we obtain
the expansion coefficients of functions ${\cal K}^1$ and ${\cal K}^2$ as
\begin{eqnarray}
K^1_{mn}
&=&
\frac{1}{m+n+1}\left[
  \left(\frac{1}{2}+2x\right)P_m^+Q_n^+
 +\left(\frac{1}{2}-2y\right)P_m^-Q_n^-\right]\,,\\
K^2_{mn}
&=&
\frac{1}{m-n-2x-2y}
 \left[
  \left(\frac{1}{2}+2x\right)P_m^+Q_n^+
 -\left(\frac{1}{2}-2y\right)P_m^-Q_n^-
 \right]\,.
\end{eqnarray}
The coefficient $K_{mn}$ in (\ref{kexp})
is the sum of these two coefficients.
We defined $P_n^\pm$ and $Q_n^\pm$ as the coefficients of
the expansions
\begin{equation}
\frac{P(u)}{1\pm u}=\sum_{n=0}^\infty P_n^\pm u^n,\quad
\frac{Q(v)}{1\pm v}=\sum_{n=0}^\infty Q_n^\pm v^n.
\end{equation}

\section{Numerical comparison of $K^{xy}$ for NS sector}
\label{numerical.app}
In this appendix, we compare the matrix $K$ (\ref{eq:K}) computed
in \S\ref{boundary.sec} with (\ref{psiktilde}) in
\S\ref{explicit.section} .
Since we cannot perform the analytic computation
of the products of infinite matrices in (\ref{eq:K}), we have to 
perform a numerical analysis.  We truncate matrices $n$ and $\tilde n$
to a size of $500\times 500$ and numerically evaluate the matrix product. 
When $\eta=1$ (or $\eta_l=\eta_r=\eta_b$), matrix $K$ 
in the second expression ($-\tilde n^{-1}(1+n)$) gives
\begin{equation}
\left(
\begin{array}{llllll}
 0. & 0.499996 & 0. & 0.124994 & 0. & 0.062493 \\
 -0.500003 & 0. & 0.624992 & 0. & 0.187489 & 0. \\
 0. & -0.625007 & 0. & 0.624989 & 0. & 0.195299 \\
 -0.125004 & 0. & -0.62501 & 0. & 0.632798 & 0. \\
 0. & -0.187509 & 0. & -0.632826 & 0. & 0.632796 \\
 -0.0625048 & 0. & -0.195324 & 0. & -0.632829 & 0.
\end{array}
\right),
\end{equation}
where only the first $6\times 6$ entries are shown.
The matrix is in good agreement with $K^{00}$,
\be
\left(
\begin{array}{llllll}
 0. & 0.5 & 0. & 0.125 & 0. & 0.0625 \\
 -0.5 & 0. & 0.625 & 0. & 0.1875 & 0. \\
 0. & -0.625 & 0. & 0.625 & 0. & 0.195313 \\
 -0.125 & 0. & -0.625 & 0. & 0.632813 & 0. \\
 0. & -0.1875 & 0. & -0.632813 & 0. & 0.632813 \\
 -0.0625 & 0. & -0.195313 & 0. & -0.632813 & 0.
\end{array}
\right)\,.
\ee

When $\eta=\eta_b \eta_l=-1$,  (\ref{eq:K}) gives
\be\label{e:K(-1)}
\left(
\begin{array}{llllll}
 0. & -1.5 & 0. & -0.875016 & 0. & -0.687531 \\
 1.502 & 0. & -0.373998 & 0. & -0.0617483 & 0. \\
 0. & 0.374998 & 0. & -0.875008 & 0. & -0.429703 \\
 0.878016 & 0. & 0.876508 & 0. & -0.491057 & 0. \\
 0. & 0.0624983 & 0. & 0.492182 & 0. & -0.773449 \\
 0.691281 & 0. & 0.431578 & 0. & 0.774855 & 0.
\end{array}
\right)\,.
\ee
This matrix should agree with $K^{-1,1}$, which is
\be\label{Kanaly}
\left(
\begin{array}{llllll}
 -2. & -1.5 & -1. & -0.875 & -0.75 & -0.6875 \\
 1.5 & 0. & -0.375 & 0. & -0.0625 & 0. \\
 -1. & 0.375 & -0.5 & -0.875 & -0.375 & -0.429688 \\
 0.875 & 0. & 0.875 & 0. & -0.492188 & 0. \\
 -0.75 & 0.0625 & -0.375 & 0.492188 & -0.28125 & -0.773438 \\
 0.6875 & 0. & 0.429688 & 0. & 0.773438 & 0.
\end{array}
\right)\,.
\ee
The agreement is limited to the components $K_{nm}$ with
$n+m=\mathrm{odd}$!  The nonvanishing components $n+m=\mathrm{even}$
originate from the oscillator insertion at the corner.  Such extra terms cancel
if we use $(K^{-1,1}+K^{1,-1})/2$, 
and then (\ref{Kanaly}) coincides with (\ref{e:K(-1)}).

In this way, we have seen that (\ref{eq:K}) gives a correct formula
only when there are no operators inserted at the corners.  If such insertion is
necessary, one should use (\ref{kexp}) derived from the correlation function.


\begin{thebibliography}{99}
\bibitem{r:review}
Some of the review articles are,\\
  P.~Di Vecchia and A.~Liccardo,
  %``D branes in string theory. I,''
  NATO Adv.\ Study Inst.\ Ser.\ C.\ Math.\ Phys.\ Sci.\  {\bf 556}, 1 (2000)
  [arXiv:hep-th/9912161],
  %%CITATION = NASCD,556,1;%%
% P.~Di Vecchia and A.~Liccardo,
%``D-branes in string theory. II,''
  arXiv:hep-th/9912275,\\
  %%CITATION = HEP-TH/9912275;%%
  M.~R.~Gaberdiel,
%  ``Lectures on non-BPS Dirichlet branes,''
  Class.\ Quant.\ Grav.\  {\bf 17}, 3483 (2000)
  [arXiv:hep-th/0005029],\\
  %%CITATION = CQGRD,17,3483;%%
V.~Schomerus,
%  ``Lectures on branes in curved backgrounds,''
  Class.\ Quant.\ Grav.\  {\bf 19}, 5781 (2002)
  [arXiv:hep-th/0209241].
  %%CITATION = CQGRD,19,5781;%%

\bibitem{IIM} 
 Y.~Imamura, H.~Isono and Y.~Matsuo,
%  ``Boundary states in open string channel and CFT near corner,''
  Prog.\ Theor.\ Phys.\  {\bf 115}, 979 (2006)
  [arXiv:hep-th/0512098].
  %%CITATION = PTPKA,115,979;%%
  
\bibitem{IM}
  H.~Isono and Y.~Matsuo,
%  ``Boundary state in open string channel and open / closed string field
%  theory,''
 "Quantum Theory and Symmetry IV" 
 Vol. 1, P229-241,
(ed. V. K. Dobrev, Heron Press, 2006), 
 [arXiv:hep-th/0511203].
  %%CITATION = HEP-TH/0511203;%%

\bibitem{Others}
  A.~Ilderton and P.~Mansfield,
  %``Timelike T-duality in the string field Schroedinger functional,''
  JHEP {\bf 0510}, 016 (2005)
  [arXiv:hep-th/0411166],\\
  %%CITATION = JHEPA,0510,016;%
  D.~Gaiotto, L.~Rastelli, A.~Sen and B.~Zwiebach,
  %``Star algebra projectors,''
  JHEP {\bf 0204}, 060 (2002)
  [arXiv:hep-th/0202151].
  %%CITATION = JHEPA,0204,060;%%

\bibitem{LPP}
A.~LeClair, M.~E.~Peskin and C.~R.~Preitschopf,
  %``String Field Theory on the Conformal Plane. 1. Kinematical Principles,''
  Nucl.\ Phys.\  B {\bf 317}, 411 (1989);
  %%CITATION = NUPHA,B317,411;%%
% A.~LeClair, M.~E.~Peskin and C.~R.~Preitschopf,
  %``String Field Theory on the Conformal Plane. 2. Generalized Gluing,''
  Nucl.\ Phys.\  B {\bf 317}, 464 (1989).
  %%CITATION = NUPHA,B317,464;%%
  
\bibitem{DFMS}
  L.~J.~Dixon, D.~Friedan, E.~J.~Martinec and S.~H.~Shenker,
  %``The Conformal Field Theory Of Orbifolds,''
  Nucl.\ Phys.\  B {\bf 282}, 13 (1987).
  %%CITATION = NUPHA,B282,13;%%

  


%\cite{Di Vecchia:1997pr}
\bibitem{DiVecchia:1997pr}
  P.~Di Vecchia, M.~Frau, I.~Pesando, S.~Sciuto, A.~Lerda and R.~Russo,
  %``Classical p-branes from boundary state,''
  Nucl.\ Phys.\  B {\bf 507} (1997) 259
  [arXiv:hep-th/9707068].
  %%CITATION = NUPHA,B507,259;%%

\bibitem{0211250}
  M. Billo, M. Frau, F. Fucito, A. Lerda, A. Liccardo and I. Pesando,
  JHEP {\bf 0302} (2003) 045,
  [arXiv:hep-th/0211250]. \\
  See also \\
  %\cite{Chen:2000ks}
  B.~Chen, H.~Itoyama, T.~Matsuo and K.~Murakami,
  %``Worldsheet and spacetime properties of p - p' system with B field and
  %noncommutative geometry,''
  Nucl.\ Phys.\  B {\bf 593} (2001) 505
  [arXiv:hep-th/0005283],
  %%CITATION = NUPHA,B593,505;%%
  %\cite{Chen:2000pk}
  % B.~Chen, H.~Itoyama, T.~Matsuo and K.~Murakami,
  %``Correspondence between noncommutative soliton and open string / D-brane
  %system via Gaussian damping factor,''
  Prog.\ Theor.\ Phys.\  {\bf 105} (2001) 853
  [arXiv:hep-th/0010066],\\
  %%CITATION = PTPKA,105,853;%%
  %\cite{Murakami:2001un}
  K.~Murakami,
  %``p-p' system with B field and projection operator noncommutative
  %solitons,''
  JHEP {\bf 0108} (2001) 042
  [arXiv:hep-th/0104243],\\
  %%CITATION = JHEPA,0108,042;%%
  which evaluate correlation functions with two twist fields, 
  which are similar to the definitions of the OBS.

\bibitem{HH}
  K.~Hashimoto and H.~Hata,
  %``D-brane and gauge invariance in closed string field theory,''
  Phys.\ Rev.\  D {\bf 56}, 5179 (1997)
  [arXiv:hep-th/9704125]\\
  %%CITATION = PHRVA,D56,5179;%%
See also recent developments, \\
  Y.~Baba, N.~Ishibashi and K.~Murakami,
  %``D-branes and closed string field theory,''
  JHEP {\bf 0605}, 029 (2006)
  [arXiv:hep-th/0603152],
  %%CITATION = JHEPA,0605,029;%%
%  Y.~Baba, N.~Ishibashi and K.~Murakami,
  %``D-brane States and Disk Amplitudes in OSp Invariant Closed String Field
  %Theory,''
  arXiv:0706.1635 [hep-th].
  %%CITATION = ARXIV:0706.1635;%%
  
\bibitem{KMW}
  I.~Kishimoto, Y.~Matsuo and E.~Watanabe,
  %``Boundary states as exact solutions of (vacuum) closed string field
  %theory,''
  Phys.\ Rev.\  D {\bf 68}, 126006 (2003)
  [arXiv:hep-th/0306189],
  %%CITATION = PHRVA,D68,126006;%%
%  I.~Kishimoto, Y.~Matsuo and E.~Watanabe,
  %``A universal nonlinear relation among boundary states in closed string
  %field theory,''
  Prog.\ Theor.\ Phys.\  {\bf 111}, 433 (2004)
  [arXiv:hep-th/0312122].\\
  %%CITATION = PTPKA,111,433;%%
  I.~Kishimoto and Y.~Matsuo,
  %``Cardy states, factorization and idempotency in closed string field
  %theory,''
  Nucl.\ Phys.\  B {\bf 707}, 3 (2005)
  [arXiv:hep-th/0409069].
  %%CITATION = NUPHA,B707,3;%%
 
\bibitem{IIM3}
Y. Imamura, H. Isono and Y. Matsuo, to appear. 
 
\bibitem{GJ3}
 D.~J.~Gross and A.~Jevicki,
  %``OPERATOR FORMULATION OF INTERACTING STRING FIELD THEORY. 3. NSR
  %SUPERSTRING,''
  Nucl.\ Phys.\  B {\bf 293}, 29 (1987).
  %%CITATION = NUPHA,B293,29;%%

\bibitem{FMS}
  D.~Friedan, E.~J.~Martinec and S.~H.~Shenker,
  %``Conformal Invariance, Supersymmetry And String Theory,''
  Nucl.\ Phys.\  B {\bf 271}, 93 (1986).
  %%CITATION = NUPHA,B271,93;%% 

\end{thebibliography}
\end{document}